\documentclass[compsoc, conference, a4paper, 10pt, times]{IEEEtran}

\usepackage{listings}
\usepackage{xcolor}
\usepackage{graphicx}
\usepackage{caption}
\usepackage{subcaption}
\usepackage{todonotes}
\usepackage{multirow}
\usepackage{booktabs}
\usepackage{xspace}
\usepackage{paralist}
\usepackage{amsmath}
\usepackage{tabulary}
\usepackage{prettyref}
\usepackage{verbatim}
\usepackage{float}
\usepackage{epsfig}
\usepackage{graphicx}
\usepackage{amssymb}
\usepackage{pifont}
\usepackage{helvet}
\usepackage[T1]{fontenc}
\usepackage[utf8]{inputenc}
\usepackage{authblk}
\usepackage{flushend}
\usepackage{times}
\usepackage{makecell}
\usepackage{url}
\usepackage{balance}

\usepackage[colorlinks=true,urlcolor=black]{hyperref}
\def\BibTeX{{\rm B\kern-.05em{\sc i\kern-.025em b}\kern-.08em
    T\kern-.1667em\lower.7ex\hbox{E}\kern-.125emX}}
\hypersetup{hidelinks}

\usepackage{pifont}

\newcounter{dn}

\lstdefinestyle{cstyle}{
  basicstyle=\footnotesize\ttfamily,
  keywordstyle=\color{black!85}\bfseries,
  keywordstyle=[2]\color{black!85}\bfseries\emph,
  showstringspaces=false,
  language={C},
  breaklines=false,
  mathescape=true,
  escapechar={@}
}
\lstdefinestyle{inline}{
  style=cstyle,
  mathescape=false,
  breaklines=true,
  keywordstyle=,            %
  keywordstyle=[2],
  extendedchars=true,
  basicstyle=\ttfamily\small
}

\lstdefinelanguage
   [x64]{Assembler}     %
   [x86masm]{Assembler} %
   {morekeywords={CDQE,CQO,CMPSQ,CMPXCHG16B,JRCXZ,LODSQ,MOVSXD, %
                  POPFQ,PUSHFQ,SCASQ,STOSQ,IRETQ,RDTSCP,SWAPGS, %
                  LFENCE,CPUID,CMP,CLFLUSH,ENDIF,IFDEF
                  }}
\newrefformat{cha}{\hyperref[#1]{Chapter~\ref*{#1}}}
\newrefformat{app}{\hyperref[#1]{Appendix~\ref*{#1}}}
\newrefformat{sec}{\hyperref[#1]{Section~\ref*{#1}}}
\newrefformat{sub}{\hyperref[#1]{Section~\ref*{#1}}}
\newrefformat{tab}{\hyperref[#1]{Table~\ref*{#1}}}
\newrefformat{fig}{\hyperref[#1]{Figure~\ref*{#1}}}
\newrefformat{line}{\hyperref[#1]{line~\ref*{#1}}}
\newrefformat{lst}{\hyperref[#1]{Listing~\ref*{#1}}}
\newrefformat{pat}{\hyperref[#1]{Patch~\ref*{#1}}}
\newrefformat{alg}{\hyperref[#1]{Algorithm~\ref*{#1}}}

\newcommand{\Code}[1]{\lstinline[style=inline,breaklines=false]@#1@}
\newcommand{\Go}{Go\xspace}
\newcommand{\SSP}{SSP\xspace}
\newcommand{\ROP}{ROP\xspace}
\newcommand{\spac}{SPEAR\xspace}%

\let\paragraph\relax
\newcommand{\paragraph}[1]{\textbf{#1}}

\begin{document}

\title{Bypassing memory safety mechanisms through speculative control flow hijacks}

 \author[1,2]{Andrea Mambretti}
 \author[1]{Alexandra Sandulescu}
 \author[1]{Alessandro Sorniotti}
 \author[2]{\\William Robertson}
 \author[2]{Engin Kirda}
 \author[1]{Anil Kurmus}
 \affil[1]{IBM Research -- Zurich, R\"uschlikon, Switzerland}
 \affil[  ]{\textit {\{asa, aso, kur\}@zurich.ibm.com}}
 \affil[2]{Northeastern University, Boston, USA}
 \affil[  ]{\textit {\{mbr, wkr, ek\}@ccs.neu.edu}}

\maketitle
\pagestyle{empty}

\begin{abstract} 
The prevalence of memory corruption bugs in the past decades resulted in 
numerous defenses, such as stack canaries, control flow integrity (CFI), and 
memory-safe languages. These defenses can prevent entire classes of 
vulnerabilities, and help increase the security posture of a program. In this 
paper, we show that memory corruption defenses can be bypassed using speculative 
execution attacks. We study the cases of stack protectors, CFI, and bounds 
checks in \Go, demonstrating under which conditions they can be bypassed by a 
form of speculative control flow hijack, relying on speculative or architectural 
overwrites of control flow data. Information is leaked by redirecting the 
speculative control flow of the victim to a gadget accessing secret data and 
acting as a side channel send. We also demonstrate, for the first time, 
that this can be achieved by stitching together multiple gadgets, in a 
speculative return-oriented programming attack. We discuss and implement 
software mitigations, showing moderate performance impact.
\end{abstract}

\section{Introduction}
\label{sec:intro}

Memory corruption vulnerabilities have plagued the computer security field for 
more than 30 years. Multiple ways of exploiting memory bugs have surfaced, 
requiring controls to be placed at different levels in the software stack: 
mechanisms such as stack canaries and control flow integrity have been 
designed and deployed as a mitigation in existing software, while new languages 
were designed with memory safety to close this class of bugs in new 
programs~\cite{memory:szekeres13,vanderveen12}. 

Recently, a new class of attacks, transient execution 
attacks~\cite{canella18arxiv}, and more specifically speculative execution 
attacks~\cite{kocher18oakland, kiriansky18specoverflow, mais18rsb, 
koruyeh18woot, schwarz:netspectre, BhattacharyyaSN19, 
mambretti19woot} have been the subject of intense scrutiny. The ensuing 
vulnerabilities appear difficult to mitigate without considerable performance 
trade-offs, leading to the conclusion that speculative execution attacks will 
remain a problem for the foreseeable future, and therefore a possibly fruitful area of 
research~\cite{mcilroy2019spectre}.

A natural question to ask is whether the advent of transient execution attacks 
has changed the security stance of modern computing systems against memory 
corruption attacks: does the security of memory safety mechanisms, such as stack 
smashing protection (SSP), control flow integrity (CFI), and those embedded in 
memory safe languages, hold in the post-Spectre threat model?

In this paper, we show that multiple memory safety mechanisms that would
otherwise successfully prevent exploitation of vulnerabilities can be
speculatively bypassed to perform arbitrary memory reads. Because these
attacks require a combination of techniques, we show that they do not apply
to all memory safety mechanisms and a careful, case-by-case analysis is necessary.

At a high level, these attacks work by overwriting, either architecturally or 
speculatively, a backwards or forward edge, followed by the use of speculative 
code reuse attacks to leak data.  In all cases, this overwrite achieves a 
speculative control flow hijack, i.e., a redirection of the speculative control 
flow to an attacker-chosen arbitrary address. One case of such an attack is the 
\emph{speculative buffer overflow} discovered by Kiriansky and 
Waldspurger~\cite{kiriansky18specoverflow}, where a return address is 
speculatively overwritten.

We demonstrate that SSP, GCC's vtable verification (VTV), and \Go's runtime
memory safety checks are all vulnerable. In particular, we develop a practical
attack against SSP, where the mitigations against a stack-based buffer overflow
in \Code{libpng} can be speculatively bypassed to read arbitrary bytes from the
victim program.  This attack additionally leverages a last level cache (LLC)
eviction attack to extend the speculative execution window, and a speculative
return-oriented programming (ROP) attack to achieve a Flush+Reload side channel
by reusing 5 gadgets from the victim program.  Both components of the attack
are not specific to SSP and generalize beyond our selected use case. Our
results demonstrate that, although such end-to-end attacks are not trivial to
mount, they are realistic. For this reason, we evaluate countermeasures for each
attack scenario, showing that mitigations are both effective and viable from a
performance standpoint.

This paper makes the following contributions:
\begin{itemize}
\item Demonstration of a practical attack against SSP-based buffer overflow
    mitigations, together with proof-of-concept attacks against GCC VTable Verification (VTV)
    and against \Go's array bounds checks.

\item Demonstration of speculation window lengthening leveraging LLC eviction of victim data.

\item Practical speculative code reuse attack (ROP) to achieve side-channel send.

\item Custom mitigations derived from \Code{lfence}-based and masking-based approaches,
    withstanding the class of speculative architectural control flow hijacking attacks,
    together with a performance evaluation.
\end{itemize}

\section{Speculative execution attacks on memory safety mechanisms}
\label{sec:spearE2EDescription}

In this section, we describe end-to-end speculative execution attacks on 
abstracted memory safety mechanisms. We begin with a high-level overview of the 
various components necessary to perform such an end-to-end attack. We then 
proceed to analyze the class of speculative control flow hijacks which is at the 
heart of the attack; we refer to this general category of attacks as SPEculative 
ARchitectural control flow hijacks, or \spac, and detail them in 
\prettyref{sec:spac_overview}. Furthermore, we analyze the eviction mechanism in 
\prettyref{sec:spac_evict}, and the speculative ROP in \prettyref{sec:spac_rop}.

\label{sec:overview}
\begin{figure}[h]
    \centering
    \includegraphics[width=\linewidth]{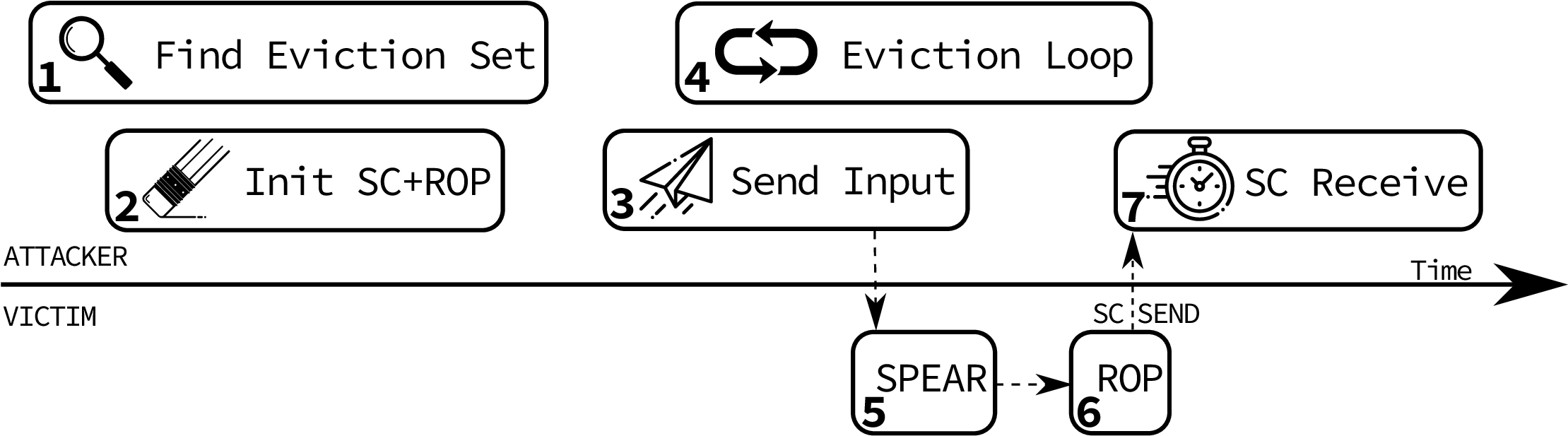}
    \caption{Overview of speculative attack against memory safety mechanisms.}
    \label{fig:init_scheme}
\end{figure}

\prettyref{fig:init_scheme} shows an overview of the steps required to perform 
an end-to-end attack. The attack has a preparation phase (Steps 1 and 2), where 
eviction sets (to ensure the existence of a suitably long speculation window) 
are identified, memory used by the side channel is flushed and 
ROP gadgets are primed in the instruction cache. The attacker then submits an 
input to the victim in Step 3, crafted to trigger a violation of a memory safety 
property. We assume that traditional exploitation of the violation is prevented 
by a suitable memory safety mechanism. However, the attacker uses a speculative 
execution attack to bypass the mechanism by overwriting (architecturally or 
speculatively) control-flow data, and obtaining a speculative control flow 
hijack (Step 5). As a result, the victim is tricked into executing a 
side-channel send of attacker-chosen memory in Step 6: this is achieved with the 
ROP component, which reuses code snippets from the victim program, appropriately 
selected and primed in the initialization phase. The attacker can then execute 
the corresponding side-channel receive in Step 7. The success rate of the attack 
is increased by concurrently executing an eviction loop to lengthen the 
speculation window (Step 4) using the eviction sets found in Step 1.

\textbf{Threat model:} The general threat model for all attacks in this paper
is a local unprivileged attacker, targeting a process holding a secret in memory. We do not
assume that the attacker is able to inject code in the victim program's address
space. We assume the attacker has knowledge of the victim program code, as well
as the virtual address of code at runtime as is the case for \Go, or that they
can recover this information if randomized, possibly by using
microarchitectural side
channels~\cite{sacham04,hund2013practical,evtyushkin2016jump}. Finally,
because we opt to use a speculative ROP payload, we assume a hyperthread-colocated attacker, thereby sharing the instruction cache with the victim program, which the attacker leverages during the ROP chain warm-up phase.
The goal of the attacker is, as in all transient execution attacks, to leak
secrets from the target program.
Attacks based on the architectural overwrite of a backward or forward edge
correspond to the case where an attacker can provoke a memory safety violation
whose traditional exploitation is prevented by hardening mechanisms in place.
This is demonstrated in the SSP and CFI use cases. In this case, we assume that
the victim program can either be executed multiple times by the attacker or
that the program automatically restarts, given that each attack run leaks
a limited volume of information and likely leads to abnormal program termination. 
This assumption remains realistic in practice because modern Linux
distributions with \Code{systemd} automatically restart services after abnormal termination.

Attacks based on the speculative overwrite of a forward edge correspond to
a victim program with a memory safety check that the attacker can exercise and
speculatively bypass. This is demonstrated in the \Go use case. In this case,
given that the overwrite of control flow data occurs speculatively, the attack
does not lead to program termination, and so the attack does not require the
ability to restart the victim.

\subsection{\spac attacks}
\label{sec:spac_overview}

A \spac-vulnerable sequence is a code sequence that results in a 
speculative control flow hijack. A speculative control flow hijack allows an 
attacker to gain control of the target program's speculatively-executed code. 
This is a powerful primitive: an attacker can follow such an attack with a 
speculative ROP sequence to speculatively execute 
code gadgets that access a secret and send it to the attacker via a side 
channel.

\begin{figure}[h]
    \centering
    \includegraphics[width=\linewidth]{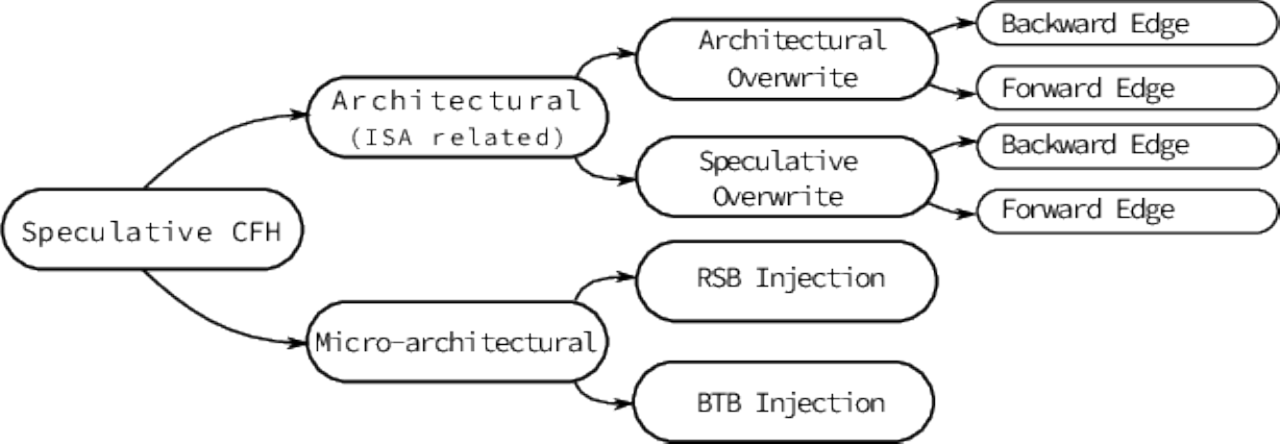}
    \caption{Overview of various Speculative control flow hijacking attacks.}
    \label{fig:categ}
\end{figure}

\prettyref{fig:categ} shows a breakdown of the various instances in the \spac 
attack class in the context of different variants of speculative control flow 
hijacks.
Classic speculative control flow hijacking attacks can be performed 
through microarchitectural components such as the Branch Target Buffer (BTB) and 
Return Stack Buffer (RSB)~\cite{kocher18oakland, mais18rsb, koruyeh18woot}. At the same 
time, the speculative control flow can also be influenced by instruction 
sequences that only affect architectural components, such as registers or 
memory: we refer to these as \spac attacks. For instance, executing the 
\Code{call \%rbx} x86 instruction speculatively when the value of \Code{\%rbx} 
is available to the execution unit will result in speculative execution continuing 
at the address in the \Code{\%rbx} register. Therefore, if the \Code{\%rbx} 
register can be controlled by the attacker, a speculative control flow hijack 
can occur. This control by the attacker can either be architectural or 
speculative, as we will see next.

Similarly, a \Code{push \%rbx; ret} instruction sequence with the register value 
available would also simply continue execution at the provided address, with no 
need to predict where speculative execution continues via the RSB. Hence, 
\spac-vulnerable code patterns can concern both forward edges (\Code{jmp} and 
\Code{call}) and backward edges (\Code{ret}).

The \spac classification offers us a convenient way to 
reason on attacks triggered by control flow data overwrite. \spac 
covers all attack scenarios studied in this paper, namely, speculative bypass of 
memory safety mechanisms; in addition, it covers other known attacks, 
such as the speculative overwrite of a backward edge~\cite{kiriansky18specoverflow}, and the speculative 
bypass of manually-inserted array bounds checks~\cite{MSSpeculativeExecutionDevGuide}.

\subsubsection{Architectural overwrite}
\label{sec:spac_arc}

The case where an attacker controls the control-flow-influencing register 
architecturally, i.e., via the Instruction Set Architecture (ISA), is closely 
related to traditional memory corruption attacks. These attacks can nowadays be
mitigated by mechanisms such as stack smashing protection (SSP) and, in 
general, CFI implementations that check the validity of control flow metadata 
before control flow is transferred, thus detecting and preventing outcomes 
induced by attacker-controlled overwrites. \spac architectural overwrite attacks 
focus on the opportunity that the attacker has to speculatively bypass the
checks introduced by these mitigations.

\begin{lstlisting}[frame=none,
                   language={[x64]Assembler},
                   caption=Architectural backward edge overwrite.,
                   basicstyle=\scriptsize\ttfamily,
                   label=lst:arc_bwd]
;Copy of RET Value
    mov rax,[rsp]
    mov QWORD[stored_ret], rax

;Architectural Overwrite
; (Attacker Controlled)
    mov rax, QWORD[target]
    mov [rsp], rax

;Evict RET Value Copy
    clflush [stored_ret]
    lfence

;Backward Edge Integrity Check
; (Speculation Trigger)
    mov rax, [rsp]
    cmp rax, QWORD[stored_ret]
    jne my_exit

;Backward Edge Hijack
    ret
\end{lstlisting}

We provide in \prettyref{lst:arc_bwd} the snippet of code that illustrate the
backward edge case for architectural overwrites. A similar snippet for the
forward edge case is reported
in~\prettyref{lst:arc_fwd},~\prettyref{app:snippets}. The structure of both
cases is similar: the original value of the edge (\prettyref{line:copy_target})
is preserved in a safe location, after which, we assume that the architectural
overwrite is performed (\prettyref{line:overwrite}) with an attacker-controlled
value (e.g., through a buffer overflow). Afterwards, the program executes an
integrity check on the forward or backward edge (\prettyref{line:intcheck})
before performing the control flow transfer (e.g., SSP or CFI check). To increase
the success rate of the attack, we try to maximize the speculation window caused
by the integrity check, for instance by evicting its reference value -- in the
snippet, this step is captured by a \Code{clflush} instruction
(\prettyref{line:evict}). If the CPU mispredicts the outcome of the check, it
might execute either a \Code{ret} (backward edge) or a \Code{call} (forward
edge) which transfers the control towards the attacker-controlled value used in
the architectural overwrite (\prettyref{line:CFH}).

\subsubsection{Speculative overwrite}
\label{sec:spac_spec}

Alternatively, the attacker may control the control-flow-influencing register 
speculatively. This means that in a first phase, speculative execution is 
triggered (for example by a conditional branch). In a second phase, the 
attacker speculatively influences the control flow edge, thus hijacking
speculative control flow. The control-flow-influencing value may be the 
result of a load from an address that is generated during the speculative 
execution phase, or it may be loaded from a location that is speculatively 
overwritten by a preceding store operation, resulting in speculative 
store-to-load forwarding.

We provide in \prettyref{lst:spec_bwd} the snippet of code that illustrate the
backward edge case for speculative overwrites. A similar snippet for the forward
edge case is reported in~\prettyref{lst:arc_fwd},~\prettyref{app:snippets}. Both
cases share the same structure. First, speculative execution is triggered by a
condition (\prettyref{line:trigger}). Then, the speculative overwrite is
performed through some instruction within the speculated part of the code. Here,
the value used for the overwrite is under attacker control
(\prettyref{line:spec_overwrite}). Finally, the overwritten value is used for
control flow transfer allowing the attacker to hijack the speculative control
flow (\prettyref{line:spec_hijack}).

\begin{table}[ht!]
\centering
\begin{tabular}{c c c c c }
    \multicolumn{1}{c}{}&\multicolumn{2}{c}{\bf{Architectural}}&\multicolumn{2}{c}{\bf{Speculative}} \\
    \Xhline{1.2pt}
    \multicolumn{1}{c}{\bf{Family}}&\multicolumn{1}{c}{\it{Fwd}}&\multicolumn{1}{c}{\it{Bwd}}&\multicolumn{1}{c}{\it{Fwd}}&\multicolumn{1}{c}{\it{Bwd}}\\
    \Xhline{1.2pt}
    \multicolumn{1}{r}{Intel Broadwell}&99.5&94.9&99.5&98.7\\
\hline
    \multicolumn{1}{r}{Intel Skylake}&97.6&98.3&98.2&92.1\\
\hline
    \multicolumn{1}{r}{Intel Coffee Lake}&99.8&98.1&99.7&99.4\\
\hline
    \multicolumn{1}{r}{Intel Kabylake}&99.5&95.9&100&99.5\\
\hline
    \multicolumn{1}{r}{AMD Ryzen}&100&100&100&100\\
\Xhline{1.2pt}
\end{tabular}
    \caption{Success rate (in percentage) computed over $1000$
    iterations for architectural or speculative
    overwrites of backward and forward edges performed on various architectures
    families.}
    \label{tab:speculator_tests}
\end{table}

\subsubsection{\spac experimental results}
\label{sec:spac_experiment}

We follow the methodology of Mambretti et al.~\cite{mambretti19acsac} and test all four snippets using 
the Speculator tool~\cite{SpeculatorGit}, which aids the detection of 
speculative control flow transfers by using performance monitor counters (PMC) and \emph{speculation markers}.

The \spac experimental results are shown in~\prettyref{tab:speculator_tests}.
Each success rate is computed on $1000$ iterations.
In the \emph{architectural overwrites} case, speculative control flow hijacks are observed at 
least $95$\% of the time for \prettyref{lst:arc_bwd} and $97$\% of the time
for \prettyref{lst:arc_fwd} on all tested architectures. The results prove
that control flow is indeed speculatively transferred to the overwritten location, 
thereby bypassing the checks during speculative execution. Therefore, we conclude 
that \spac attacks with architectural overwrites can result in speculative 
control flow hijacks.
In the \emph{speculative overwrites} case, for the backward edge case the 
success rate is at least $92$\% while for the forward edge case it is at least
$98$\% .
The experiment results demonstrate that speculative overwrites are feasible
and lead to speculative control flow hijacks provided a sufficiently large
speculation window exists to facilitate the edge overwrite followed by the
dereference.

\begin{lstlisting}[frame=none,
                   language={[x64]Assembler},
                   caption=Speculative backward edge overwrite.,
                   basicstyle=\scriptsize\ttfamily,
                   label=lst:spec_bwd,escapechar=|]
;Speculative execution trigger
    ...|\label{line:trigger}|

;Speculative Overwrite
; (Attacker Controlled)
    mov rax, QWORD[target]
    mov [rsp], rax   |\label{line:spec_overwrite}|

;Backward Edge Hijack
    ret                   |\label{line:spec_hijack}|
\end{lstlisting}

\subsection{Speculation window and eviction}
\label{sec:spac_evict}

\spac attacks require the existence of a 
speculation window to permit the execution of the control flow transfer and the 
side channel send operation, a common precondition for all speculative execution 
attacks. This requires a speculative execution trigger, i.e., an instruction 
that causes a wide-enough window of dependent instructions that are executed but 
not retired. This is usually 
achieved when the process accesses uncached data: the speculation window then 
corresponds to the time for the access to main memory to complete. In 
\prettyref{lst:arc_fwd} for example, this is achieved with the \Code{clflush} 
instruction. To show the necessity of a wide speculation window, we re-run the snippet without \Code{clflush} in the 
Speculator tool and verify that indeed the control flow hijack only takes place 
in about one run out of $1000$. When it does, the window is only a couple of 
instructions wide. We therefore conclude that without eviction, or other similar 
approaches to lengthen the speculation window, \spac attacks are unlikely to be 
practical.

In all snippets referenced by this section, the speculation window is 
artificially lengthened by flushing one of the memory operands of the compare 
instruction. This may not be realistic, as it imposes a strong requirement on 
the victim code to include a flush (or comparable) instruction. Instead, because 
the last level cache (LLC) is shared and often inclusive, the same effect can be 
accomplished more realistically by an external attacker thread computing an 
\emph{eviction set} and performing a small number of accesses to addresses in 
this set.
An LLC eviction set competes for the same LLC slice and cache set as the target 
address to be evicted. Existing techniques for performing such attacks typically 
assume knowledge of the targeted physical address, as the LLC is physically 
indexed. As a consequence of rowhammer attacks, this is no longer realistic, as 
most OSes have removed access to physical mappings for unprivileged users.
In Linux, privileged-only \Code{/proc/PID/pagemap} access~\cite{linux-pagemap} was
introduced in release 4.0.

We demonstrate here that such eviction attacks can still be performed without 
knowledge of the physical address. To this end, we perform the eviction in two 
steps. The first step consists of the identification of an eviction set for a 
cache line in a page under the attacker's control, by following the approach of 
Maurice et al.~\cite{MauriceSNHF15}. The second step consists in releasing this page to 
the OS, and executing the victim process such that it reuses the
previously-created page. This permits the reuse of the eviction set constructed and 
verified to be working in the first step. To increase the victim data eviction success
rate, we follow the eviction set loading method proposed by Liu et al.~\cite{liu15sp}.
We show details of such a practical attack in 
\prettyref{sec:canaryEviction} for SSP. 

\subsection{Speculative ROP}
\label{sec:spac_rop}

To perform a complete speculative execution attack, the speculative control flow 
hijack must be followed by a \emph{side channel send} gadget with a secret 
input. Unfortunately, Spectre~v1-type Flush+Reload side channel send gadgets are known to be 
difficult to find~\cite{kocher18oakland, projectzero:spectre}. As in classical 
control flow hijacks~\cite{roemer12rop}, however, a speculative code reuse attack 
can be performed by chaining the speculative execution of gadgets
to construct a Flush+Reload side channel send sequence.
To chain the gadget sequences, we proceed in a similar way to traditional ROP attacks, with sequences ending in \Code{ret} instructions, yet with two additional requirements. 
These requirements for performing speculative code reuse are the following: 
\begin{inparaenum}[\itshape i\upshape)]
	\item execution of all instructions in the gadgets must fit into the speculation window;
	\item all code pages in which the gadgets reside must be present
		and mapped in the victim process.
\end{inparaenum}

The first requirement is a consequence of the behavior of speculative execution.
In particular, all return values used to chain gadgets need either to be in store buffers or in cache. 
Indeed, whether the return addresses are speculatively or architecturally written to the stack, execution of return instructions will make use of these addresses if they are available, with the CPU preferring those values for steering front-end fetches over values provided by the RSB. If the return address is not in cache (or in store buffers), loading the return address from memory will exceed the speculation window in practice and only RSB-based branch prediction will be in use, which will result in failure of the attack.
A similar approach and analogy exists with forward edges for code reuse.
Using the Speculator tool, we obtain experimentally that the maximum number of empty gadgets that fit
in the largest speculation window is 20 on our Kaby Lake i7-8550 test platform.

The second requirement is needed to avoid page misses during gadgets execution.
In the event of a page miss, speculative execution might halt or nested
speculation might be triggered. Despite of the two strict requirements, we show in
\prettyref{sec:SSPSpeculativeROP} that speculative ROP can be achieved for a
practical use case.

\section{Case studies}

In this section, we analyze different case studies where memory safety
mechanisms can be bypassed with \spac attacks. In particular, in
\prettyref{sec:ssp} we use a practical attack that speculatively bypasses SSP
leveraging architectural overwrites of backward edges. \prettyref{sec:cfi}
analyzes architectural overwrites of forward edges, targeting two prominent CFI
frameworks, GCC VTV and LLVM CFI. In the GCC VTV case, we show how the
integrity check of the forward edge can be used to perform a speculative
control flow hijack. For LLVM CFI, we conclude that the constraints of its
implementation does not allow SPEAR attacks to be mounted in practice,
demonstrating the importance of careful feasibility analysis. Finally, in
\prettyref{sec:goAndFriends}, we demonstrate two types of speculative bounds
check bypasses in the \Go language using speculative overwrites of a forward
edge. We show how the attacker may influence the control flow target through
both a load whose address value is attacker controlled and a load of a value
that was speculatively overwritten by the attacker. We demonstrate practical
implementation of speculative ROP and LLC eviction techniques as part of the
end-to-end practical attack on SSP, i.e., we implement all the stages
in~\prettyref{fig:init_scheme}. We do not further re-implement them in the
case of CFI and \Go, where they would equally apply and where we focus instead on the
central part of the attack as a proof of their feasibility, i.e., we implement only
Step 5 in~\prettyref{fig:init_scheme}. Therefore, the success rates reported
below refer either to all the stages together for the SSP case ($7.19$\%) or
just the hijack stage for the Go (above $80$\%) and the GCC VTV ($85$\%) use
cases, hence the large difference.

\subsection{Attacking stack canaries} \label{sec:ssp}

Stack canaries are one of the earliest mitigations against buffer 
overflows~\cite{memory:szekeres13}, and are widely used to this day. Among the 
most broadly adopted implementations are LLVM's and GCC's Stack Smashing 
Protection (SSP) and Microsoft's \Code{/GS}. At a high level, stack canaries 
work by inserting a value (the \emph{canary}) between stack buffers and 
control-flow influencing data on the stack, in particular the saved return 
value. The integrity of the canary is then checked prior to using the saved 
return value. Local stack variables are reordered such that buffers, likely to 
be overflowed, reside adjacent to the canary while code pointers remain further 
away. This way, contiguous overflows of local stack buffers can be detected by 
the integrity check. The chosen canary value is randomly generated once during 
process execution start, and stored in a safe location. 

Each compiler performs the instrumentation differently but in essence the 
mechanics are identical with respect to \spac attacks; we therefore focus on the 
example of LLVM on Linux \Code{x86_64}. Implementations consist of two distinct 
\emph{instrumentation atoms}. The instrumentation atoms on our target system are 
shown in \prettyref{lst:ssp-check}. The first, the prologue \SSP atom, is placed 
after the function prologue and local variable allocation, and is responsible 
for storing the canary value on the current stack frame. The second, the 
epilogue \SSP atom, is placed before local variable deallocation and the 
function epilogue. It compares for equality  between the global and local canary 
values; if the values differ, the \Code{__stack_chk_fail} function is called, 
terminating the program. If the local canary value was not modified during 
function execution, the function returns normally. We show next that this 
particular comparison can be the target of a \spac attack.

\subsubsection{\spac attack on LLVM-SSP}

The pattern of the \SSP instrumentation closely resembles that of 
\prettyref{lst:arc_bwd}. Under our threat model, an attacker with a buffer 
overflow against a function protected by SSP can perform a \spac architectural 
overwrite attack of the return value of that function. We describe a practical
attack targeting a version of \Code{libpng} with a reported 
buffer overflow (CVE-2004-0597): the bug is not exploitable in the traditional 
way owing to the fact that the function is compiled with SSP. We show how a 
speculative adversary can exploit the \spac architectural overwrite to leak 
arbitrary secrets from the victim.

The attack proceeds as follows: in the first step, the attacker overwrites the 
saved return address of the victim function. In the second step, the attacker leverages a misprediction in the 
conditional jump of the canary integrity check, thus transiently executing a 
return to the previously overwritten return address. This PHT-based 
misprediction is achieved by the attacker in a way similar to Spectre~v1, by 
executing the canary integrity check with an intact local canary sufficiently 
many times. As discussed in \prettyref{sec:spac_evict}, another requirement is 
that a sufficiently long speculation window exist. We achieve this by evicting 
the global canary from the LLC, as we show in \prettyref{sec:canaryEviction}. 
The attacker is then able to perform a side-channel send operation by 
constructing a speculative ROP chain to access a secret as we show
in \prettyref{sec:SSPSpeculativeROP}.

\begin{lstlisting}[frame=none,
                   language={[x64]Assembler},
                   caption=Stack canary check instrumentation example.,
                   basicstyle=\scriptsize\ttfamily,
                   label=lst:ssp-check]
func:
    prologue

; Store canary on the stack
    mov rbx, QWORD[fs:0x28]
    mov QWORD[stack_canary], rbx
    ...
    body
    ...
; Check for corrupted canary, if yes fail
    mov rbx, QWORD[stack_canary]
    xor rbx, QWORD[fs:0x28]
    je exit
    call __stack_chk_fail
exit:
    epilogue
    ret
\end{lstlisting}

\subsubsection{LLC eviction of the global canary}
\label{sec:canaryEviction}

We apply the two-step method described in \prettyref{sec:spac_evict} for the
eviction of the global canary from LLC, and thus from all cache levels by the
property of inclusiveness of caches on the target platform. The global canary
value is always stored at a fixed offset in a page: we use this property to
find eviction sets for this particular offset by using the undocumented Intel
LLC slice function reverse engineered by Maurice et al.~\cite{MauriceSNHF15}.

The attacker process first identifies a page with a known eviction set and then
unmaps it to be reused by the victim to store its canary. This is achieved with
two processes under attacker control, as follows. At first, one of them maps a
hugepage and enters a loop in which it brings an eviction set into cache and
waits for feedback from the second attacker process. The latter in turn probes
its own stack canary and reports back a success as soon as the canary is no
longer cached. 
Once the eviction set is identified, the attacker releases the page, which is
now ready to be reused by the victim process to store its canary. The page
release is done via \Code{madvise} which instructs the system about the process
memory behavior, in this case indicating that a certain memory range will not
be accessed soon (\Code{MADV_DONTNEED}).  We manually craft the memory area
released by the attacker in order to shift the target page frame in the right
position in the kernel buddy freelist. We empirically verify that the reuse of
a page frame for the victim canary page occurs with $100$\% success rate when
attacker controls victim startup. When the attacker does not control victim
startup, the success rate drops (but remains above $50$\%), because
synchronization is more difficult and all processes in the system consume
resources from the buddy freelist. Factors that influence the success rate
include the order of the page in the freelist, the ``distance'' between the
release operation by the attacker and the request operation by the victim
process. We note that we do not use any artificial synchronization mechanism
between the victim and attacker, which makes this attack widely applicable.

While data eviction is a common part for speculative execution attacks, we
adapt an LLC eviction technique only used previously in the context of side
channels. Existing techniques for Spectre attacks evict large quantities of
data from the caches, lowering success rate. For the SSP attack, this technique
ensures the ROP gadgets executed speculatively remain in cache. Their eviction
would result in the attack failing, because the RSB would be used to predict
return location.

\subsubsection{Speculative ROP}
\label{sec:SSPSpeculativeROP}

We now focus on building and using a speculative ROP chain that accesses a 
secret and leaks it through a side channel. We use the Flush+Reload cache 
side channel initially used by Kocher et al.~\cite{kocher18oakland}, although 
other side channels can be used similarly~\cite{BhattacharyyaSN19, 
schwarz:netspectre, mambretti19woot}.

In \prettyref{sec:spac_rop}, we have identified two major constraints on the
attack:
\begin{inparaenum}[\itshape i\upshape)]
	\item a limited number of instructions can fit into the speculation window;
        and,
	\item all code pages in which the gadgets reside must be present and mapped with corresponding TLB entries.
\end{inparaenum}
In addition to these requirements, we note that gadget code, as well as any data
accessed by gadgets, must be available in cache during speculative execution.
Typically, this is not an issue in speculative execution attacks because the
attacker can run several attack iterations as warm up phase to
bring the required data in the cache, whereas this attack is
\emph{single shot}: the process terminates after each attempt and this 
is an additional requirement.

\begin{lstlisting}[frame=none,
                   language={[x64]Assembler},
                   basicstyle=\scriptsize\ttfamily,
                   caption=Example of Flush+Reload gadget.,
                   label=lst:specv1]
mov rax, secret
shl rax, 8
add rax, shared_array
mov rax, [rax]
\end{lstlisting}

Concerning the first requirement, the Flush+Reload side-channel send gadget only 
requires a few instructions: there are sufficiently short gadgets available, and 
length is therefore not an issue in practice. For the second requirement 
instead, we create a tool to search for gadgets in code that was recently 
accessed by the victim program, for which pages are present and mapped in the 
victim process. The tool traces the victim process and collects all executed 
shared (library) code pages, which are then fed into an existing ROP gadget 
search tool, ROPgadget~\cite{ropgadget}. 
We run the tool on the victim program and find 26 mapped code pages within the
4 different modules used by the victim: \Code{libc}, \Code{libpng}, \Code{libz}
and \Code{ld}. In total, the tool discovered 2096 gadgets, out of which 406 are
candidates for building the side-channel send gadget. Per-gadget occurrences are
shown in \prettyref{tab:rop_occ}.
Finally, to ensure that all gadget sequences are in cache, a 
hyperthread-colocated attacker performs a \ROP chain warm up phase by executing 
the chain in close temporal proximity with the \spac attack.

\begin{table}[ht!]
\centering
\begin{tabular}{c c}
    \Xhline{1.2pt}
    \multicolumn{1}{c}{\bf{Gadget type}}&\multicolumn{1}{c}{\it{Occurrence}}\\
    \Xhline{1.2pt}
    \multicolumn{1}{r}{pop reg ; ret}&262\\
\hline
    \multicolumn{1}{r}{mov reg1, [reg0] ; ret}&69\\
\hline
    \multicolumn{1}{r}{shl reg, 8 ; ret}&4\\
\hline
    \multicolumn{1}{r}{add reg1, reg0 ; ret}&71\\
\Xhline{1.2pt}
\end{tabular} \caption{ROP gadgets used for building Spectre v1 chain with their
  corresponding occurrences.  The search space is a subset of \Code{libc},
  \Code{libpng}, \Code{libz} and \Code{ld} executable pages, obtained by
  filtering out pages unmapped in the victim's address space and pages without
  a valid TLB mapping.}
    \label{tab:rop_occ}
\end{table}

We build a 5-gadget ROP chain using the \ROP gadgets found by our gadget
search tool. The chain is functionally equivalent to the 
Flush+Reload gadget shown in \prettyref{lst:specv1}. The chain accesses a 
target address computed using a secret byte value, as in the initial Spectre 
attacks~\cite{kocher18oakland}. Because Flush+Reload requires shared memory, we 
choose the target address to reside in such a shared memory area between 
attacker and victim, the first 16 readable and executable pages of the 
\Code{libpthread} library. To leak one byte we use an array size of 256. To avoid 
prefetching effects during side-channel receive, we choose the element size to 
be 256, i.e., four cache lines. The total array size equals \Code{256 x 256} 
bytes, 16 pages.

By splitting the Flush+Reload gadget in small
sequences of instructions as shown in \prettyref{lst:specv1}, we easily find
the required gadgets within the constraints of the 
attack. The ROP chain that we find and use in the attack is shown 
in \prettyref{lst:spadgets}. This chain pops the addresses (controlled by the attacker) of the start 
of the 16 pages and of the targeted secret from the stack. Next, the secret value 
is loaded at \prettyref{line:load}. The next speculative gadgets multiply the secret
value by 256 and compute the target address. The last speculative gadget
dereferences the target address, resulting in a load being issued during speculative
execution. This eventually brings the value into the cache to be observed by the attacker.
The whole chain therefore allows the attacker to implement a universal read
primitive over the victim process speculatively, using a Flush+Reload attack and 
the attacker's control over the stack.

\begin{minipage}{\linewidth}
    \begin{lstlisting}[frame=none,
                       language={[x64]Assembler},
                       basicstyle=\scriptsize\ttfamily,
                       caption=Flush+Reload gadget ROP chain.,
                       label=lst:spadgets,
                       escapechar=|]
libpng.so.3.1.2.5 : 0x7960
    pop rdx
    ret
libpng.so.3.1.2.5: 0x7f0a
    pop rsi
    ret
libpng.so.3.1.2.5 : 0x128ec
    mov eax, dword ptr [rsi] | \label{line:load} |
    mov byte ptr [rdi + 6a], al
    ret
libpng.so.3.1.2.5 : 0x9f4b
    shl rax, 8
    add rax, rdx
    ret
libpng.so.3.1.2.5 : 0x9fde
    add eax, dword ptr [rax]
    add byte ptr [rdi], cl
    xchg eax, ebp
    ret
\end{lstlisting}
\end{minipage}

\subsubsection{Attack evaluation and results}
\label{sec:sspAttackEvalRes}

The attacker targets the \Code{libpng} version 1.2.5 which is vulnerable to 
\Code{CVE-2004-0597}~\cite{libpngcve}.

\Code{CVE-2004-0597} is a stack buffer overflow which allows the attacker
to read \Code{length} bytes in \Code{readbuf}. Due to improper sanitization
of \Code{length}, a read larger than \Code{PNG_MAX_PALETTE_LENGTH} is allowed
in a stack buffer. The target victim is a program that receives a \Code{.png}
file and parses the file using the unpatched \Code{libpng-1.2.5}. When building the victim target with stack
canaries enabled, the compiler will instrument \Code{png_handle_tRNS} with the
corresponding prologue and epilogue \SSP atoms. As expected, \SSP
protects \Code{png_handle_tRNS} from exploitation by stopping execution
before the function returns. However, using a \spac architectural overwrite attack, we can perform a 
speculative control flow hijack. During the \spac attack, the attacker feeds 
\Code{.png} files of the legitimate length to train the pattern history table
to bypass the stack canary check. Then, the attacker provides a \Code{length}
larger than \Code{PNG_MAX_PALETTE_LENGTH} that overwrites the value of the
return address to trigger the speculative ROP attack.

We confirm the attack works and leaks bytes at arbitrary, attacker-chosen
addresses from the victim memory, on Intel Skylake and Coffee Lake with latest
microcode updates, and on Ubuntu 16.04 and 18.04 (both with kernel version
4.15.0) with all default Spectre mitigations enabled. Namely, both setups include
\lstinline[style=inline,breaklines=true]{__user pointer sanitization} and
\lstinline[style=inline,breaklines=true]{usercopy/swapgs barriers}
mitigations against \Code{Spectre v1}. Moreover, default mitigations
against \Code{Spectre v2} are present (\Code{retpoline}, \Code{IBPB},
\Code{IBRS_FW}, and \Code{RSB filling}), excepting \Code{STIBP} which is disabled on Ubunutu 16.04.
We report the attack evaluation results on Intel(R) Core(TM) i7-6700K CPU @ 4.00GHz (Skylake)
running Ubuntu 16.04.6 with kernel version 4.15.0. As described 
in \prettyref{sec:spearE2EDescription}, the attack has an initialization phase 
where eviction sets are identified, memory used by the side channel is flushed and
the ROP sequence is primed. Concurrently with the submission of the 
malicious payload, the attacker also runs the eviction loop to lengthen the 
speculation window by causing the eviction of the stack canary in the victim.

\begin{lstlisting}[frame=none,
                   language=C,
                   caption=\Code{libpng} vulnerable snippet related to
                   \Code{CVE-2004-0597}.,
                   basicstyle=\scriptsize\ttfamily,
                   label=lst:libpng-cve,
                   escapechar=|]
void /* PRIVATE */
png_handle_tRNS(png_structp png_ptr, png_infop info_ptr, png_uint_32 length)
{
  ...
  png_byte readbuf[PNG_MAX_PALETTE_LENGTH];
  ...
  if (png_ptr->color_type == PNG_COLOR_TYPE_PALETTE) {
    if (!(png_ptr->mode & PNG_HAVE_PLTE))
    {
     /* Should be an error, but we can cope with it */
     png_warning(png_ptr, "Missing PLTE before tRNS");
    }
    ...
    png_crc_read(png_ptr, readbuf, (png_size_t)length);
    png_ptr->num_trans = (png_uint_16)length;
  }
  ...
}
\end{lstlisting}

We measure the attack success rate as the number of times the attacker is able
to correctly guess a secret byte from the victim memory space, per total number
of runs.
We report means over $100$ runs with $95\%$ confidence level. The
end-to-end attack success rate is $7.19\% \pm 0.62$, for a single run. In
practice the attacker does, as in most other Spectre attacks, re-run the attack
as many times as necessary to improve its guesses and reach close to $100$\%
success rate. Therefore, we
compute the leakage rate based on the attack time, which is measured as the
full duration of repeating the attack 100 times against the re-startable
victim. The duration includes the restart time of the victim and the attacker
execution time. The end-to-end leakage rate of victim bytes is $0.3$ bytes per
second (with all correct guesses), which we deem sufficiently high for
practical use. Due to different \Code{binutils} versions in the two
distribution versions, we observe a slight leakage rate drop in the Ubuntu
18.04 environment.

For improving the success rate, and therefore improving the
leakage rate of the end-to-end SSP attack, one needs to improve the success of
each individual stage of the attack showed in~\prettyref{fig:init_scheme}. In
addition, the attacker may run too fast or too slow with respect to the victim
(the attacker simply attempts to synchronize with busy loops), which can also
lead to failure of the attack. We have verified such synchronization is
successful in our PoC in $78$\% of cases. We already report in \prettyref{sec:canaryEviction} 
results for the LLC eviction stage: $100$\%.
Because changes to the victim can affect the success rate, measuring the success of other
individual steps within the end-to-end PoC is very difficult. 
However, based on these numbers and experiments outside the PoC, we infer that the
greatest area for improvement in the leakage rate should come from improving
the ROP gadget phase (e.g., limiting cases where the gadget code is not in
cache) and side channel receive/send (e.g., limiting cache noise from eviction
activity and other sources, or using another side channel).

\subsection{Attacking CFI}
\label{sec:cfi}

Control Flow Integrity (CFI) of forward edges aims to protect the integrity of 
code pointers used in indirect calls and jumps. CFI implementations contain two 
main parts: instrumenting all indirect control transfers to check their validity 
at runtime, and classifying valid control flow transfers (typically using static 
analysis at build time). We analyze here two prominent cases: the GCC Virtual 
Table Verification (VTV)~\cite{tice2014enforcing} mechanism to prevent \Code{c++} virtual 
table corruption, as well as LLVM-CFI~\cite{kuznetsov14}, a 
publicly available, low overhead, forward-edge CFI implementation. In the GCC 
VTV, we prove that a \spac attack is possible, while in the LLVM-CFI case we 
conclude that eviction-related considerations result in the speculation window 
being too short for practical exploitation. In particular, this case study 
demonstrates that we cannot conclude that \spac attacks apply equally to all 
implementations of memory safety-related defenses, and case-by-case analysis is 
necessary.

\subsubsection{GCC VTV}
\label{sec:gcc-vtv}

In the GCC VTV implementation, for every call to a virtual function in the 
program, the compiler inserts a check to make sure that the pointer used for the 
indirect call belongs to the virtual table of the object. Such check is 
represented by a call to the function \Code{__VLTVerifyVtablePointer} 
implemented in \Code{libvtv.so} library. Within this function, the pointer is 
looked up from the table; if found, the function simply returns to the program 
which will perform the call, otherwise, it gracefully fails. If an attacker can 
successfully evict the cache line related to the variable the pointer is tested 
against, speculative execution is triggered during the evaluation of the check. 
In that case, the indirect call to the virtual function is speculatively 
executed and the code at the corrupted pointer is executed. At this point, the 
attacker has performed speculative control flow hijack and can mount a data 
exfiltration attack as described in \prettyref{sec:ssp}.

In our proof-of-concept implementation of this attack, we artificially evict 
from all cache levels the variable related to the vtable of the object within 
the \Code{libvtv.so} code. Then, we create a \Code{c++} program that defines two 
different classes each containing one virtual method. The first class is our 
target for the forward edge overwrite. To verify whether speculative 
control flow hijack takes place, we instrument the program to read performance 
monitor counters and set the speculative control flow hijack target to 
contain a \textit{speculation marker}. We use the second class to instantiate 
the object that is later corrupted. 

After object initialization, we perform a vtable pointer overwrite in our victim 
object making it point to the vtable of the first class. Finally, we perform the 
virtual call for the control flow transfer which is instrumented by GCC VTV with 
a call to the integrity check inside the \Code{libvtv.so} library. During normal 
execution, this overwrite is detected by the library which reports the 
corruption and prevents the control flow transfer by terminating the 
application. With a \spac attack as described here, we verify that control flow 
hijacking occurs in ~$85$\% ($n{=}1000$), demonstrating that a \spac architectural 
forward-edge attack is viable against GCC VTV. We note also that the redirection 
is performed to a vtable of a completely unrelated class, a case which should be 
prevented by VTV. A real-world attack would additionally require evicting the 
compare variable, for example by using the same method as in 
\prettyref{sec:canaryEviction}, as well as a way of achieving a side-channel 
send for the attacker, as in \prettyref{sec:SSPSpeculativeROP}. 

\subsubsection{LLVM CFI}
\label{sec:llvm-cfi}

The CFI solution implemented in LLVM uses function types as \emph{equivalence 
classes}: an indirect call to a function of a different type than the one 
specified by the programmer is forbidden by the CFI instrumentation. This is 
achieved by placing functions of an equivalence class in a jump table, thereby 
having as many jump tables (whose addresses are carefully chosen) as equivalence 
classes. The instrumentation for indirect calls then consist in simply checking 
that the address of the target fall within the range of the jump table, and at 
the right alignment.

This range check can be seen as a check against a compile-provided constant 
value, using the address of the provided target. Both of these components are by 
design available and cached while performing this check: evicting the code that 
contains the range check would result in speculative execution stopping, and 
evicting the address of the target would result in the iBTB being used for 
speculative execution. In either case, a \spac attack would fail. The attack may 
be triggered without any attempt to artificially extend the speculation window, 
but, as demonstrated experimentally in \prettyref{sec:spac_evict}, the resulting 
speculation window is rare and short, making such attacks unlikely to be 
practical. We conclude that LLVM CFI is in practice not vulnerable to \spac 
attacks. 

\subsection{Attacking memory safe languages}
\label{sec:goAndFriends}

Most modern languages are designed to ensure memory safety. Instrumental to 
achieving this property are bounds checks for load and store operations into 
arrays. In this section, we show how bounds checks may be speculatively bypassed, 
allowing the transient execution of out-of-bounds load and store operations. We 
show under which conditions this leads to a \spac attack.  

We focus in this case study on the popular \Go programming language, runtime 
and compiler. We present two variants, one where data that influences a forward 
control flow edge is architecturally overwritten and one where a forward edge is 
speculatively overwritten. In either case, the attacker is able to achieve a 
speculative control flow hijack. We prototype both variants and show the
conditions under which the attack succeeds at a rate exceeding $80$\%.

\begin{lstlisting}[frame=none,
                   caption=Arrays in \Go.,
                   basicstyle=\scriptsize\ttfamily,
                   label=lst:go_slicestruct]
type slice struct {
    array unsafe.Pointer
    len   int
    cap   int
}
\end{lstlisting}

Before detailing the two attacks, we give a brief introduction to the way the 
\Go compiler manages arrays and bounds checks. Arrays in \Go are represented in 
memory as the \Code{struct} shown in \prettyref{lst:go_slicestruct}. The address 
of the contiguous chunk of virtual memory backing the array is stored in \Code{array}. 
The number of elements that \Code{array} can hold (and 
implicitly the size of the memory chunk since \Go is statically typed and the size 
of the elements is always known) is stored in \Code{cap}. The current number of elements that 
have been stored in the array is stored in \Code{len}.

Whenever an array access is performed in \Go, the compiler will add appropriate 
bounds checks. This is achieved in the course of the compiler pass to translate 
the abstract syntax tree (AST) into the static single assignment (SSA) intermediate 
representation by adding an \Code{IsInBounds} meta-operation before every array 
load or store. \Code{IsInBounds} takes two arguments, the index of the current 
access and the length of the array, and drives a conditional jump either to the 
basic block that performs the array access if the index is between zero and 
length minus one, or a jump to a function that raises a \Code{panic} otherwise.

\begin{lstlisting}[frame=none,
                   language={[x64]Assembler},
                   basicstyle=\scriptsize\ttfamily,
                   caption=Bounds check in \Go.,
                   label=lst:go_boundsCheckAsm]
mov rcx, [array]
cmp [array+0x8], rax
jbe runtime.panicindex
mov rax, [rcx+rax*8]
\end{lstlisting}

\Code{IsInBounds} is translated by later passes into a sequence of instructions 
similar to the one shown in \prettyref{lst:go_boundsCheckAsm}. The snippet shows 
a load from an array of integers: at first \Code{rcx} is loaded with the address 
of the memory array, a compare instruction is issued between the index of the 
array access in \Code{rax} and the array length at \Code{array+0x8}. If the 
index is negative or not strictly less than the length, the code jumps to a call 
to the \Code{runtime.panicindex} function. Otherwise the array access is 
performed.

The conditional jump generated by the \Code{IsInBounds} meta-operation may 
speculatively execute the wrong jump target and perform a transient load or 
store operation out of bounds. We show two distinct code patterns, one 
leveraging a load and one a store, that may lead to speculative control flow 
hijack.

\begin{lstlisting}[frame=none,
                   numbers=none,
                   basicstyle=\scriptsize\ttfamily,
                   caption=Load-based speculative control flow hijack code
                   pattern.,
                   label=lst:go_golangPoc1]
array[index].function()
\end{lstlisting}

\subsubsection{Load-based \spac speculative attack}
\label{sec:goPoC1}

The first pattern is shown in \prettyref{lst:go_golangPoc1}.  It represents an
instance of a \spac-speculative attack and consists of an interface function
call, where the interface is stored into an array of interfaces \Code{array},
dereferenced at position \Code{index}. Note that the array must be an array of
interfaces so that calling the function is achieved by an indirect call. For the
attack to be successful, we need \Code{index} to be attacker-controlled and the attacker must be able to store the value of two pointers in the memory space of the target process at a known location.
The first condition is met whenever a process accesses an array using an index
that is received as an external input. The second condition is very commonly met
since programs store user-provided input for processing. Knowledge of the
location of the stored pointers depends on the memory area being used, and is
aided by the deterministic nature of the \Go allocator.

\begin{lstlisting}[frame=none,
                   caption=Structs used by interface calls.,
                   basicstyle=\scriptsize\ttfamily,
                   label=lst:go_itab]
type iface struct {
    tab  *itab
    data unsafe.Pointer
}

type itab struct {
    inter *interfacetype
    _type *_type
    hash  uint32
    _     [4]byte
    fun   [1]uintptr
}
\end{lstlisting}

Without loss of generality, we describe the case where \Code{function} is the 
first function defined by the interface. Exploitation proceeds as follows: 
first, the attacker prepares the memory structures that are used when an 
interface call is performed. The structures are shown 
in \prettyref{lst:go_itab}, and are used by dereferencing the \Code{tab} pointer 
from the \Code{iface} struct and then calling into the \Code{fun} array. 

\begin{lstlisting}[frame=none,
                   caption=Memory layout in preparation for the exploitation of load-based speculative control flow hijack. The attacker fake \Code{iface} starts at offset 0x0. The fake \Code{itab} prepared by the attacker starts at offset 0x1000. The control flow hijack target is located at offset 0x2000.,
                   numbers=none,
                   language=C,
                   basicstyle=\scriptsize\ttfamily,
                   label=lst:go_memlayout]
fake iface:
0x0000:         <fake itab>
0x0008:  0x0000000000000000
             ...
fake itab:
0x1000:  0x0000000000000000
0x1008:  0x0000000000000000
0x1010:  0x0000000000000000
0x1018:        <CFH target>
             ...
CFH target:
0x2000:     <attacker code>
\end{lstlisting}

In preparation for exploitation, the attacker ensures that the memory layout 
of the target program contains a pattern similar to that shown 
in \prettyref{lst:go_memlayout}. Assuming that the attacker wants to 
speculatively redirect the control flow to address \Code{0x2000}, the attacker 
creates a fake \Code{itab} structure (in the example at \Code{0x1000}) such 
that the first entry in the \Code{fun} pointer array points to the desired 
target. Then the attacker creates a fake \Code{iface} structure (in the 
example at offset \Code{0x0}) such that the \Code{tab} pointer points to the 
aforementioned \Code{itab} structure. With the memory thus prepared, the 
attacker supplies the index into the array such that the resulting address 
(the base address plus index multiplied by the size of an \Code{iface} 
structure) equals the fake \Code{iface} structure (\Code{0x0} in our 
example). With the index thus set the program will call the 
\Code{runtime.panicindex} function; however if the conditional jump of the 
bounds check is mispredicted, the dereference and subsequent indirect call will 
take place transiently. Note that, contrary to the case studies 
in \prettyref{sec:ssp} and \prettyref{sec:cfi}, the attack is not necessarily 
``single shot'': if the program calls \Code{recover}, the attacker might be able 
to execute the vulnerable sequence multiple times.

We prototype the attack to evaluate its effectiveness in a proof of concept. The 
proof of concept only aims to establish the feasibility of the attack: in 
particular we do not integrate into an end-to-end attack and refer 
to \prettyref{sec:sspAttackEvalRes} for cache eviction and speculative ROP. The 
PoC contains the pattern of \prettyref{lst:go_golangPoc1} called in a loop to 
train the pattern history table and ensure that the bounds check conditional 
jump as strongly non-taken. The index used to access the array in the loop is in 
bounds during the training phase and is then set to the target index computed as 
described above in the last iteration.

To verify whether speculative control flow hijack takes place, we instrument 
the program to read PMCs during the execution of 
the loop, and set the speculative control flow hijack target to contain a 
speculation marker. The \Code{runtime.panicindex} function is modified 
to read and persist PMC values for each execution.

This instrumentation permits us to verify that speculative control flow hijack 
indeed takes place. The success rate is influenced by several factors that we 
review here. The most relevant factor is the size of the speculation window, 
which is influenced by how quickly the correct jump target is determined. The 
speculation window is maximized if the variables used in the compare instruction 
that drives the jump -- especially the array length -- are not present in any of 
the levels of the cache. In order to get empirical evidence of this fact, we 
instrument the program with a \Code{clflush} instruction right before the array 
dereference to ensure that the array length is not cached. In practice, an 
attacker may achieve the same result by performing cache eviction code 
sequences. However flushing the cache alone does not ensure a high success rate: 
this is because the array length is stored right after the base address of the 
array, whose address is loaded into memory as the first instruction of the 
dereference sequence. We verify that if the two memory locations belong to 
different cache lines, the speculation window is maximized. Another factor that 
influences the success rate is whether the target of the speculative control 
flow hijack is already in the instruction cache. We make sure that this 
be the case by insert a call to the marker function in the warm up phase 
before the loop. We report success rates exceeding $80\%$ ($n{=}1000$) when the array length 
is flushed and is in a separate cache line as the base address on multiple 
platforms (Xeon CPU E5-2640, Core i7-8650U, Core i7-6700K) and different 
versions of the \Go runtime (1.13.4, 1.12, 1.10.4).

\subsubsection{Store-based \spac speculative attack}

The second pattern is shown in \prettyref{lst:go_golangPoc2}. 

\begin{lstlisting}[frame=none,
                   basicstyle=\scriptsize\ttfamily,
                   numbers=none,
                   caption=Store-based speculative control flow hijack code
                   pattern.,
                   label=lst:go_golangPoc2]
array[index] = value
...
interface.function()
\end{lstlisting}

The pattern consists of a store operation of an attacker-controlled value at an 
attacker-controlled location into an array. The elements stored in the array 
must permit storage of a pointer. Smaller sizes would permit partial control 
over the speculative control flow hijack target. The pattern requires that the 
array store be followed by an interface call. The interface call does not need 
to be related to the array. It only needs to be in close proximity of the store 
operation so that it may still be speculatively executed. This pattern does not 
require any ability to perform preparatory store operations in the memory space 
of the target program. The pattern makes use of store-to-load forwarding, since 
the store in the array is used to (speculatively) overwrite a function pointer 
which is later (speculatively) loaded and called. This corresponds to the 
``speculative overwrite of forward edge'' variant of a \spac attack.

The store part of the pattern consists of a speculative version of a 
``write-what-where'' condition. It may be exploited in several ways to hijack 
the interface call: the most basic one would be to overwrite the \Code{tab} 
pointer in the \Code{iface} struct (see \prettyref{lst:go_itab}). However this 
would either require the attacker to perform a set of preparatory stores 
identical to those discussed in \prettyref{sec:goPoC1}, or it would restrict the 
freedom of the attacker to choose a target out of the existing interface 
pointers. Another strategy would be for the attacker to overwrite the \Code{fun} 
pointer in the \Code{itab} structure directly. These structures are stored in a 
non-writable virtual memory region. However, given that the store takes place 
speculatively, the attacker is able to bypass the write restrictions and 
overwrite the pointer. Therefore, we choose to prototype this simpler and more 
effective variant.

Exploitation proceeds as follows: at first the attacker speculatively overwrites 
the \Code{fun} pointer in the \Code{itab} of the interface that is later dereferenced. 
This is achieved, as the attacker controls value 
and index. The former is set to the address of the desired speculative control 
flow hijack target; the latter is set such that base array and index multiplied 
by the size of the array elements add up to the address of the \Code{fun} 
pointer to be overwritten. As in the previous section, with the index thus set 
the program will panic; however if the bounds check is mispredicted, the 
store-to-load forwarding and subsequent indirect call will take place, achieving 
speculative control flow hijack.

We prototype the attack to evaluate its effectiveness employing a similar 
instrumentation as the previous section, with PMCs and speculation markers 
employed to identify successful runs, and a loop to set the predictor state. The 
success rate is similarly influenced by ensuring that the variables driving the 
conditional branch are not cached, and that the speculative control flow hijack 
target is in cache. Under these conditions, we report success rates exceeding 
$80\%$ ($n{=}1000$) on the same platforms listed in~\prettyref{sec:goPoC1}.

\section{Mitigations}
In this section, we implement and analyze serializing-based (lfence) and
masking-based mitigations for \spac-architectural attacks (SSP)
in~\prettyref{sec:ssp_mit} and \spac-speculative ones (Go)
in~\prettyref{sec:go_mit}. We show that in both cases the masking-based solution
results in a low overhead. Finally, we discuss possible mitigations for GCC VTV
case in~\prettyref{sec:vtv_mit}.

\subsection{Mitigations for SSP}\label{sec:ssp_mit}

We investigate two possible mitigations for the \spac-architectural attack 
against \SSP. A serializing instruction such as \Code{lfence} can be inserted 
after loading the canary in the epilogue instrumentation, thereby ensuring that 
the comparison can only lead to a short enough speculation window. 
Alternatively, the return value can be masked architecturally with a generated 
value that is set to \Code{0} when the check fails (the canary is corrupted), 
and all ones when it passes, as shown in \prettyref{lst:masking-ssp}.

\begin{lstlisting}[frame=none,
                   language={[x64]Assembler},
                   basicstyle=\scriptsize\ttfamily,
                   caption=Masking mitigation sequence; \Code{rax} contains
                   global canary value and \Code{rcx} contains the stack canary;
                   \Code{rsp + 8} points to the return address.,
                   label=lst:masking-ssp]
mov rax, QWORD[fs:0x28]
mov rcx, QWORD[stack_canary]
xor rdx, rdx
cmp rax, rcx
setne dl
add rdx, 0xffffffffffffffff
and QWORD[rsp + 8], rdx
\end{lstlisting}

We implement both mitigations as compiler passes in \Code{clang+llvm}.
The masking-based mitigation implementation is an extension of
Speculative Load Hardening~\cite{SLH}.
\SSP is architecture specific, therefore our solution is built for \Code{x86_64} Linux 
systems. We run the \SSP mitigations benchmarking on Intel(R) Core(TM)
i7-6700K CPU @ 4.00GHz.
We measure the normalized runtime of both \emph{return address masking} and 
\Code{lfence} on SPECint CPU 2006. The normalized
runtime is computed as runtime over the baseline runtime constituted by 
execution with \SSP Disabled. For reference, we additionally plot the
normalized runtime for all existing \SSP implementations, \SSP Loose
(\Code{-fstack-protector} flag), \SSP Strong (\Code{-fstack-protector-strong}
flag), and \SSP All (\Code{-fstack-protector-all} flag).

\begin{figure*}
\centering
\begin{subfigure}{.5\textwidth}
	\includegraphics[width=0.97\textwidth,clip]{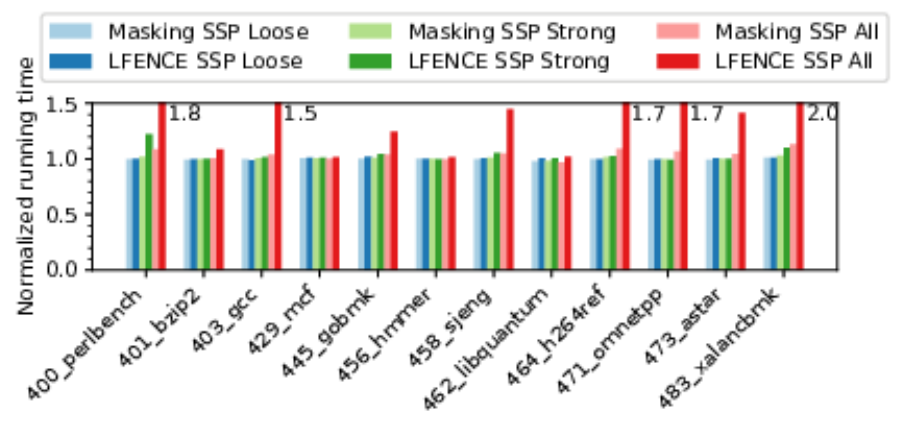}
	\centering
	\caption{\SSP with speculative bypass mitigations.}
	\label{fig:ssp-mitigations}
\end{subfigure}%
\begin{subfigure}{.5\textwidth}
	\includegraphics[width=0.97\textwidth]{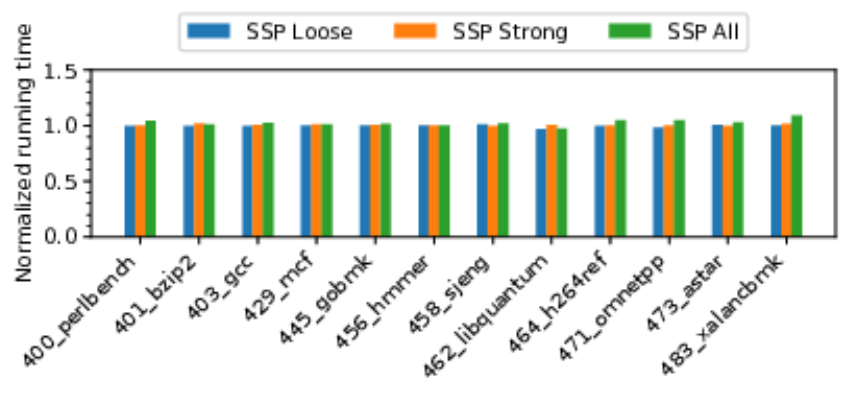}
	\centering
	\caption{Vanilla \SSP.}
	\label{fig:ssp-benchmarks}
\end{subfigure}
\caption{Overhead computed as normalized runtime over \SSP Disabled baseline.}
\label{fig:test}
\end{figure*}

The results are shown in \prettyref{fig:ssp-mitigations}. The \Code{lfence}
mitigation shows a high overhead in 9 out of 12 benchmarks, the
highest being $100$\%, in the \SSP All case with \Code{xalancbmk}.
Return address masking incurs a significantly lower, albeit still not negligible performance penalty,
reaching a maximum of $13$\% for the same benchmark. 

Based on this evaluation, we find the return address masking mitigation to be
viable and superior to the \Code{lfence} mitigation: the overhead of vanilla \SSP
(shown in \prettyref{fig:ssp-benchmarks} on SPECint CPU 2006 is at most $9$\%, in
the case of \SSP All on \Code{xalancbmk}).  In addition, we note that most Linux
distributions either use the \SSP Loose or \SSP Strong options, both of which
incur a low overhead on all \SSP benchmarks: we record a maximum of $2.1$\%
overhead over the \SSP Disabled baseline. With return address masking, the
maximum overhead becomes $2.7$\% over the \SSP Disabled baseline. We conclude that return
address masking does not impose a significant overhead with the most commonly
used \SSP compiler options.

\subsection{Mitigations for the \Go compiler}
\label{sec:go_mit}

We investigate possible mitigations for the \spac-speculative attack on \Go. The 
mitigations consist of two different compiler passes that ensure that the 
vulnerability is no longer exploitable. The first is based on \Code{lfence}, 
whereas the second is based on branchless index masking sequences. As part of 
responsible disclosure we have notified the \Go team, who have implemented 2 
families of compiler-based mitigations for Spectre, namely, index masking 
(through the \Code{-spectre=index} compiler switch) and retpoline (through the 
\Code{-spectre=ret} compiler switch). 

The first mitigation consists of adding an \Code{lfence} instruction after the 
\Code{cmp} instruction in the sequence that implements the \Code{IsInBounds} 
meta-operation. With reference to \prettyref{lst:go_boundsCheckAsm}, the 
\Code{lfence} instruction is inserted after the \Code{cmp} on line 2. The 
insertion ensures that all prior instructions have completed, which means that 
there will be no misprediction of the branch target and any out-of-bound access 
will result in a panic with no transient execution. The instruction is added 
explicitly in the pass that translates the AST into SSA form by defining a new 
\Code{Lfence} meta-operation and adding it after each \Code{IsInBounds} 
operation. We ensure that the operation is neither reordered nor eliminated.

The second mitigation we investigate entails the addition of an appropriate 
masking sequence that ensures that the index is set to a ``safe'' value in case 
of out-of-bounds accesses. The masking sequence amounts to a no-op in case the 
access is in bounds by performing an \Code{and} operation on the index with a 
sign extended $-1$ mask. If the access is not in bounds, in our implementation, 
the masking operation forces an access of the element at index $0$ in the array 
by performing an \Code{and} operation on the index with a $0$ mask. We can see 
the masking sequence in \prettyref{lst:go-go-masking}: after the usual 
\Code{cmp} and \Code{jmp} instructions, length and index are subtracted in order 
to set the carry flag. Then, the \Code{sbb} instruction is used to set a register 
to $-1$ in case of an in-bounds access or $0$ otherwise. The array is 
subsequently accessed after performing an \Code{and} operation on the index with 
the mask thus obtained. The pattern might be further optimized by using the 
\Code{cmp} instruction of the bounds check to set the carry flag. This, however, 
is not always possible since the compiler will use a compare instruction with an 
immediate whenever possible. The immediate can only be the second source 
operand, forcing the direction of the comparison instruction. For the sake of 
simplicity we therefore rely on an extra subtraction operation. The masking 
instruction sequence is added by defining three new meta-operations -- 
\Code{OpMaskStep1}, \Code{OpMaskStep2} and \Code{OpMaskStep3} -- which are later 
lowered into a \Code{sub}, \Code{sbb} and \Code{and} instruction, respectively.

\begin{figure}[ht!]
    \includegraphics[width=0.45\textwidth,trim={1.5cm 6cm 2cm 6cm},clip]{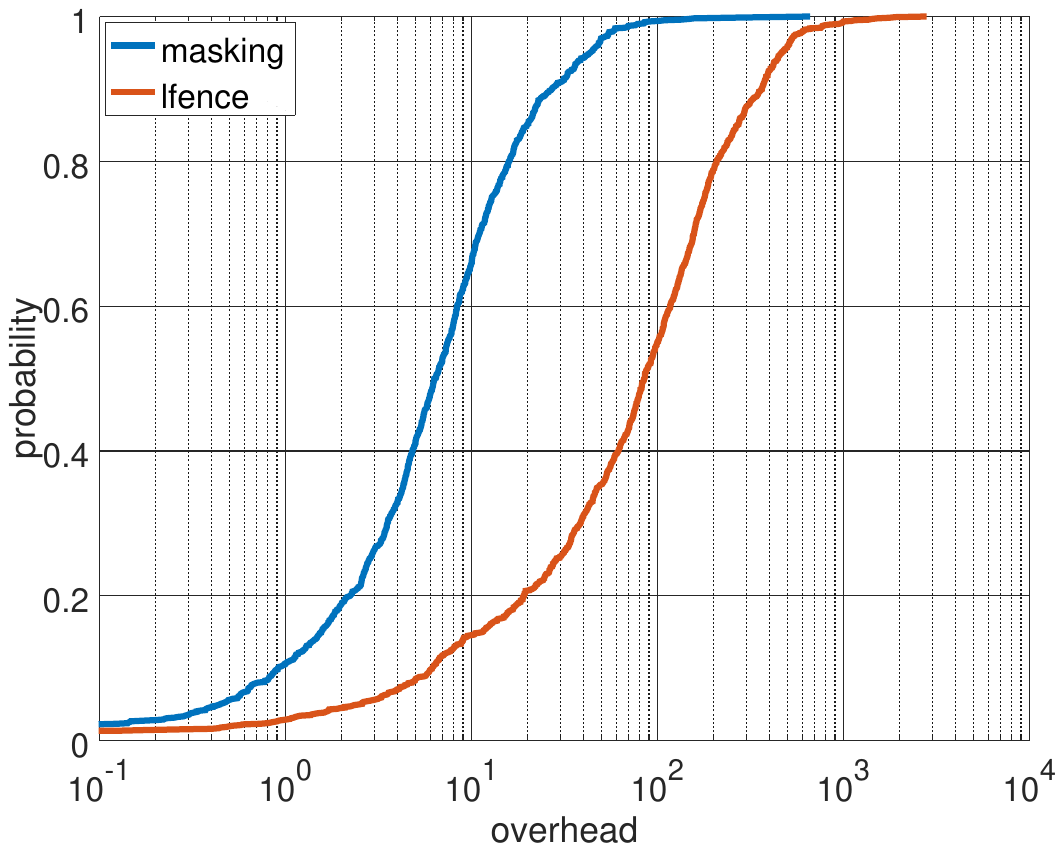}
    \centering
    \caption{Empirical CDF of the logarithm of the overhead percentage for the 
considered mitigations. Overhead data is gathered by running the full set of 
benchmarks of the \Go runtime version 1.12.0.}
    \label{fig:go-mitigations}
\end{figure}

We measure the overhead of both mitigations by building the \Go runtime version 
1.12.0 and running the full benchmark suite. We run the experiments on a 40-core 
Xeon E5-2640 machine with 64 GiB of RAM. \prettyref{fig:go-mitigations} displays 
the empirical cumulative distribution function of the overhead of each of the 
two mitigation strategies. We can see how the \Code{lfence}-based approach 
incurs a high overhead ($143\%$ mean and $84\%$ median) due to the fact that 
\Code{lfence} will terminate any speculative execution and thus severely curtail 
the instruction throughput. On the other hand, the masking approach shows a much 
lighter overhead ($12\%$ mean and $6\%$ median) since the instructions involved 
are simple and do not cause any memory-related operation.

\begin{lstlisting}[frame=none,
                   language={[x64]Assembler},
                   basicstyle=\scriptsize\ttfamily,
                   caption=Masking
  mitigation sequence; \Code{rdx} contains the index and \Code{rcx} contains
  the length of the array and \Code{rax} contains the base address of the
  array.,
                    label=lst:go-go-masking]
cmp    rcx, rdx
jae    <raise-panic>
mov    rbx, rdx
sub    rdx, rcx
sbb    rcx, rcx
and    rcx, rbx
shl    rcx, 0x4
mov    rax, [rax+rcx*1]
\end{lstlisting}

\subsection{Mitigations for GCC VTV} \label{sec:vtv_mit}
The same mitigations considered in~\prettyref{sec:ssp_mit}
and~\prettyref{sec:go_mit} work in the GCC VTV use case. Serializing mechanisms
(e.g., lfence) are a viable solution, albeit likely with high overhead. A
branchless masking solution or retpoline could also be used in this context for
what we expect to have better performance, however we did not implement these.

We believe a better approach, from a performance point of view, for GCC
VTV would be a re-design with the principles observed for LLVM CFI described
in~\prettyref{sec:llvm-cfi} where the metadata and the pointer that have to be
verified co-exist within the same cache line. This condition prevents the
attacker to achieve the correct data eviction and, consequently, the
speculation window to perform the attack is too small.

\section{Related work}
\subsection{Speculative execution attacks}
Transient execution attacks can be subdivided into two main categories:
fault-based and speculation-based attacks~\cite{canella18arxiv}. The
speculation-based, or Spectre-family, attacks comprise those leveraging
microarchitectural components such as the Pattern History Table (PHT) for
Spectre v1~\cite{kocher18oakland}, the Branch Target Buffer for Spectre v2, the Return Stack Buffer
(RSB) for Ret2Spec~\cite{mais18rsb} and Spectre returns~\cite{koruyeh18woot}.
Both BTB and RSB attacks are cases of speculative control flow hijacks,
i.e., they provide the ability for an attacker to steer speculative execution
to an arbitrary location. Varied and powerful attacks leveraging the BTB for
speculative control flow hijacks have been demonstrated, in combination with
port contention-based, instruction cache-based, or BTB-based side channels~\cite{mambretti19woot,BhattacharyyaSN19}.
In contrast, this paper focuses on \spac attacks, where the speculative control flow hijack step is based on architecturally visible control-flow
influencing instructions. In Spectre v1.1~\cite{kiriansky18specoverflow}, Kiriansky and Waldspurger
point out that speculative overwrites of backward edges lead to speculative control flow
hijacks. The \spac class includes their finding and complements it with three new types,
including the architectural overwrite case. Also, we demonstrate practical use cases
on Go memory safety and GCC VTV, and a full working attack on SSP.

In practice, BTB gadgets are hard to find, thus attacks have only been shown to
be practical if the gadget is injected (e.g., by loading attacker-controlled
eBPF bytecode into the kernel). In the \spac attacks, we reuse gadgets existing
within the victim program.  The idea of chaining speculative gadgets in a way
similar to ROP was suggested shortly after the first publication of Spectre
attacks. Some publications have referred to the same
idea~\cite{kiriansky18specoverflow, mambretti19woot}, the former only briefly
mentioning speculative ROP attacks but practical aspects are neither discussed
nor experimented on.  In contrast, this paper presents a 
practical case of chaining multiple speculative gadgets to form a cache side-channel send
gadget. 

Netspectre~\cite{schwarz:netspectre} introduces a victim data eviction
technique based on coarse-grained cache eviction. The method,
\emph{Thrash+Reload}, is a remote variant of
\emph{Evict+Reload}~\cite{gruss:cache-attacks}. The attacker starts a large
file download from the victim via a network interface. On the victim's side,
this action results in victim data eviction with a probability which depends on
the file size. \emph{Thrash+Reload} applicability is limited to scenarios where
cache thrashing does not compromise the attack: it would for instance be
detrimental in our SSP end-to-end exploit where the availability of ROP gadget
sequences in the icache is a necessary precondition. 

\subsection{Speculative execution terminology}

Canella et al. present a thorough taxonomy and evaluation of speculative
execution variants~\cite{canella18arxiv}.  In this terminology, the SSP attack
falls in the Spectre-PHT, \textit{same address-space, in-place} category, given
that the attacker triggers the to-be mispredicted victim path (successful
canary check) prior to the attack, to force the target program to self-train.
This terminology, however, does not help distinguish the different types of
speculative control flow hijacking attacks, provided in our categorization in
\prettyref{fig:categ}. In the general sense, a Spectre-PHT type misprediction
is not required for \spac attacks: other types of misprediction leading to the
overwrite of a control-flow influencing data (\spac-speculative) or other types
of misprediction following the overwrite of control-flow influencing data
(\spac-architectural) are also concerned, which justifies the need for our
categorization. For instance, \spac-speculative attack instances, such as the
Go attacks here, can be classified under the existing Spectre-PHT writing
out-of-bounds category. However, the SSP attack cannot be covered in that
category, given that no misprediction of a bounds check occurs and the write is
not speculative.

Most Spectre attacks including \spac rely on a covert channel and/or gadget to
achieve information leak. Intel terminology~\cite{intel:terminology} refers to
it as \textit{disclosure gadget}. According to their taxonomy based on
disclosure gadget location, \spac falls under the \textit{cross-domain
transient execution attack} category. While useful, this categorization also does not help
distinguish between different speculative control flow hijacks. 

\subsection{Concurrent work}
Three recent papers are concurrent to this work and partially relate to it.
Goktas et al.~\cite{goktas_speculative} demonstrate that speculative execution
attacks can be used to bypass randomization based defenses including ASLR.
Their main assumption is the presence of a powerful memory corruption
vulnerability, allowing the attacker to overwrite (architecturally) function
pointers. In contrast, SPEAR does neither aim to bypass randomization-based
defenses, nor does it assume that the attacker has access to such a powerful
vulnerability. For instance, in \prettyref{sec:ssp} we assume a vulnerability
which cannot be traditionally exploited due to the presence of SSP.
Bhattacharyya et al.~\cite{bhattacharyya20specrop} demonstrate that speculative
ROP chains can be mounted in speculative execution attacks by carefully
training the BTB (or RSB) to chain multiple speculative ROP gadgets. In
contrast, we do not make use of BTB or RSB training to chain gadgets but simply
use store-to-load forwarding of the return value on the stack. Finally, Van
Bulck et al.~\cite{vanbulck2020lvi} demonstrate chaining \Code{pop-ret}
instructions in a transient ROP attack triggered by a LVI attack. This is
complementary to SPEAR, which explores ROP in the context of speculative
execution attacks.

\subsection{Mitigations}

Since the first speculative execution attacks have been disclosed in early 2018, different
mitigations have been proposed to prevent each variant. Some mitigations are introduced at hardware
level meanwhile others are software-based. Many of these mitigations target Spectre v2 type of
attacks, meanwhile no software-transparent mitigation has been introduced for Spectre v1.

The available software-based Spectre v1 mitigations consist in either deploying a serializing instruction (e.g
\Code{lfence}) around each sensitive bounds check or, alternatively, masking the index used for accessing
arrays~\cite{kochermasking,SLH,kiriansky18specoverflow,MaskingKernel}.

While \Code{lfence} is an effective mitigation, it incurs huge performance
penalties if widely applied. Static analysis tools have been proposed to search
for sensitive code patterns. One example is the Linux kernel where vulnerable
code is instrumented on a case by case basis either through manual audit or
automatic tools (e.g., smatch~\cite{smatch}) detection~\cite{kernelmitigation}.
The drawback of current available tools is that they target Spectre v1 code
patterns such as array-out-of-bounds cases only and therefore are not useful in
the general memory corruption case (where an overwrite of a control-flow
influencing value can occur for any other mispeculation). 

At the hardware level, SpecShield~\cite{specshield} changes 
microarchitectural handling of loads and prevents forwarding of sensitive data to
probable covert channels during transient execution. It proposes three strategies to delay load
broadcast to dependent instruction until sensitive load instructions are at the top of the re-order
buffer. They demonstrate these techniques can improve performance compared to software barriers.

For Spectre v2 instead, there are software and hardware mitigations. The software mitigation
currently available is Retpoline~\cite{turner:retpoline}. This mitigation targets indirect calls and
indirect jumps and prevents them from being speculatively executed by trapping speculation within a loop. As in
the barrier cases for Spectre v1,
Retpoline requires code modification and therefore each program has to be recompiled to enforce such
mechanism. 

On the hardware side, Intel published three major protections:
\begin{inparaenum}[\itshape i\upshape)]
\item IBRS~\cite{ibrs}, which prevents speculation of indirect branches using
    target values computed using lower privileged predictor modes,
\item STIBP~\cite{stibp}, which prevents BTB poisoning from sibling threads, and
\item IBPB~\cite{ibpb}, which ensures that code before a barrier does not influence the behavior of the code after.
\end{inparaenum}
IBRS and IBPB are meant to protect higher privileged code
from lower privileged code. The only mitigation that
provides protection within the same privilege
level is STIBP, which is not enabled by default for performance reasons.
None of these Spectre v2 prevention mechanisms apply to \spac attacks, given that SPEAR does not use branch target injection.

Finally, Intel announced as part of its Control Flow Enforcement (CET)
extension, the future introduction of a new mitigation that will constrain the
target of near indirect jumps and calls to only \Code{ENDBRANCH} instructions.
Based on the release specifications, these constraints should also apply during
speculative execution. Therefore, this mitigation reduces the number of
possible gadgets where speculative execution can be redirected to during branch
target injection attacks. For \spac attacks, this mitigation applies for the
forward edge overwrite case, where it should restrict possible speculative
control flow hijack targets. For the backward edge case, Intel has implemented
a shadow stack which, if adequately enforced during speculative execution,
should stop all \spac backward edge overwrites.

\subsection{Safe speculation designs}
In addition to mitigations that aim to protect already
existing systems, several new design proposals have been presented for future architectures to prevent speculative execution attacks.

A line of research concentrates on analyzing the data flow within the
CPU pipeline and preventing unsafe operations from leaving observable effects upon misprediction. 
NDA~\cite{micro2019nda} restricts speculative data
propagation that follows an unresolved branch (potential control flow misprediction) or unresolved store address (potential memory dependence misprediction).
STT~\cite{micro2019stt} selectively forward secrets based on a speculative
taint tracking system. Dolma~\cite{usenix2019dolma} presents a lightweight
speculative information flow scheme with secure performance optimizations.
All these designs should prevent SPEAR attacks.

Another set of work, instead, proposes new cache designs.
InvisiSpec~\cite{micro2018invisispec} removes cache covert and side channels by
confining Unsafe Speculative Loads (USL) into a speculative buffer until
the USL is considered safe and the changes can be exposed to the cache
hierarchy. In a similar fashion, CleanupSpec~\cite{micro2019cleanup} prevents
the cache side-effects, however, its strategy differs from InvisiSpec because it
allows the USL to modify the cache. CleanupSpec applies an Undo
operation only when misprediction is detected, therefore limiting performance overhead.
Conditional Speculation~\cite{hpca2019condspec} and Sakalis et
al.~\cite{isca2019sakalis} block during speculation memory accesses that do not hit
the L1 cache, as the L1 accesses are safe. Finally,
DAWG~\cite{micro2018dawg} proposes a mechanism to partition the caches into
domains to provide isolation. These cache based defenses stop SPEAR attacks
as described in this paper but they do not cover cases of SPEAR where non-cache
side channels are used, such as BTB-based~\cite{mambretti19woot} or
port contention-based~\cite{BhattacharyyaSN19}.

\section{Discussion}

 \paragraph{Applicability to other use cases.} 
Beyond the highlighted use-cases, \spac attacks may be employed against other 
targets. For example, other memory-safe languages may be targeted with \spac 
attacks to speculatively bypass bounds checks as we show for the \Go 
programming language. Preliminary investigation suggests that this is likely to be 
possible, since instruction sequences for bounds checks similar to those 
detailed in \prettyref{sec:goAndFriends} are also present in \Code{Rust} and 
\Code{Java} (for JITted blocks). We analyze in more detail the Rust use
case and report our findings in~\prettyref{app:rust}.

Theoretically, any security check that directly or indirectly gates a control
flow transfer may be turned into a SPEAR attack. For instance, all the heap
hardening mechanisms that verify the integrity of the heap metadata and
pointers within \Code{libc} can potentially lead to one of the SPEAR variant
through the speculative use of a corrupted data to decide the application control
flow. However, as demonstrated in the LLVM CFI case, a case-by-case analysis is
necessary to establish whether \spac attacks are applicable.

\paragraph{Data leaked in \spac-architectural attacks.} 
\spac attacks allow an adversary to leak sensitive information from the victim
address space. In the case of SSP, we demonstrate that arbitrary memory can be
leaked, one byte per iteration. While we can target any memory location, we
cannot target data that is not deterministic across runs. In particular, we
cannot target to leak the stack canary, given that its value is re-randomized
at every program start. We note that \spac-speculative attacks do not have
this constraint, given that they do not require a program restart.

\paragraph{General applicability of speculative ROP.} 
The speculative ROP and LLC eviction techniques are
demonstrated as part of the SSP, \spac-architectural overwrite of a backward
edge, use case. Nevertheless these techniques are generally applicable for the
exploitation of other \spac use cases, with exploitability always depending on
the scenario at hand.  For the general forward edge cases, we note that this
requires, as in classical ROP attacks, a technique known as a stack pivot,
which consists in the attacker setting up a fake return stack somewhere under
its control in memory, and having the first control flow hijack point to an
instruction setting the stack pointer to that address (for instance, the
\Code{push rax; pop rsp; ret} stack pivot gadget). Using the Speculator tool, we verify
that such stack pivots do work for \spac-architectural as well
as \spac-speculative attacks.

\paragraph{General applicability of LLC eviction.} 
In our end-to-end attack over SSP, we employ a new more precise LLC eviction 
technique which is described in details in~\ref{sec:canaryEviction}. The 
necessity for developing our own, more precise, LLC eviction technique stems 
from the fact that our attack poses two additional requirements. The first is 
the fact that we require the eviction process to be very selective, since we 
cannot allow elements such as the addresses injected on the stack or the gadgets 
code to be evicted because that will stall speculative execution and prevent the 
completion of the attack. The second is that the eviction process needs to 
complete within a short amount of time to avoid the scenario where the line 
containing the canary is first evicted and then re-cached by the natural 
execution of the victim while the eviction process completes. With our 
technique, we can keep the number of possible cache-sets as small as possible 
and therefore minimize the length of the eviction process.
We explore an existing LLC flush method discovered by Oren et al.~\cite{yoren:spy}
which could potentially fit the second requirement. However, we conclude that this method is
too intrusive in a setting where the attacker relies on cached data and code
(victim secret, ROP gadgets) available in the speculation window.

\section{Conclusion} In this paper, we investigate variants of speculative
control flow hijacking attacks, called \spac, that exploit and bypass current
mitigations against classic memory corruption vulnerabilities to leak
information from local processes. With \spac, we show that Spectre-like
vulnerabilities drastically increase attack vectors for local attackers.
Therefore, they force not only the creation of new mitigations but also the
re-design of previously deployed protections. In this work, we present attacks
against stack canaries, CFI and memory-safe languages. We demonstrate a
practical attack against SSP buffer overflow mitigations and proof-of-concept
implementations against GCC VTV and \Go's runtime. We show the use of multiple ROP gadgets and
details on how to use LLC eviction without knowledge of physical addresses in
the context of \spac attacks. Finally, we discuss how \spac attacks can be
mitigated and report our performance results.

\section*{Disclosure}
We submitted the PoC exploits and our findings to the Go security team on 
November 22nd, 2019. As a result of our notification, the Go security team has 
deployed hardening measures (index masking and retpoline) which were released in 
Go 1.15.

\section*{Acknowledgement}
We would like to thank Russ Cox, Matthias Neugschwandtner, the anonymous
reviewers, and our shepherd for their valuable comments on an earlier draft of
this paper.

This work was partially-supported by National Science Foundation under grant
CNS-1703454, and ONR under the ``In Situ Malware'' project.

\balance
\label{sec:ccl}
\bibliographystyle{IEEEtranS}
\bibliography{bib/mc_sea}

\newpage
\begin{appendices}
\section{\spac attack against Rust bounds checking}
\label{app:rust}

The implementation of Rust panicking mechanism is abundant of \spac speculative
control flow hijacking patterns similar to those discussed in the \Go case
study (\prettyref{sec:goAndFriends}). Here, we examine the safety
features employed by Rust for index expressions and demonstrate a proof of
concept \spac attack against out of range access hardening.

In Rust, memory safety for index expressions is established during Mid-level
Intermediate Representation (MIR) building, with static and dynamic arrays,
slices and strings being subject to sanitization. At compiler level, index
expressions are instrumented with bounds checks which prevent out of range
access. However, similarly to the case of \Go, CPU misprediction of bounds
check outcome leads to speculative out of bounds access.
\begin{lstlisting}[frame=none,language={C},
  caption=\spac speculative control flow hijacking target in Rust.
  The \Code{index} value is attacker controlled. We assume that the attacker
  writes the CFH target in memory prior to the attack.,
  basicstyle=\scriptsize\ttfamily,
  label=lst:rust_target,
  escapechar=|]
  const PADDING_SIZE: usize = 7;
  pub type Fptr = fn(u64) -> u64;

  pub struct Data {
    _padding: [u64; PADDING_SIZE],  |\label{line:rustasm_padding}|
    buf: Box<[Fptr]>,
  }
  let data: Data = Box::new(Data { ... }); |\label{line:rustasm_box}|
  data.buf[index](); |\label{line:rustpoc_call}|
\end{lstlisting}

\begin{lstlisting}[frame=none,language={[x64]Assembler},
  caption=Disassembly of Rust index expression bounds check instrumentation.,
  basicstyle=\scriptsize\ttfamily,
  label=lst:rust_instr,
  escapechar=|]
    mov rsi, [index]
    mov rax, [buf len]
    cmp rsi, rax |\label{line:rustasm_cmp}|
    jle ok   |\label{line:rustasm_jmp}|
    call <core::panicking::panic_bounds_check>
 ok:
    mov rcx, [buf]
    mov rdx, [index] |\label{line:rustasm_deref}|
; Calls function pointer when index is in bounds
    call QWORD[rcx+rdx*8] |\label{line:rustasm_call}|
\end{lstlisting}

The attack targets the array index access followed by an indirect call in
\prettyref{lst:rust_target} at \prettyref{line:rustpoc_call}.
To trigger the panicking system, the array is accessed with an attacker
controlled index which is out of bounds. Rust MIR instruments the array index access with a bounds
check.  We analyze the index expression bounds check instrumentation at Assembly level in
\prettyref{lst:rust_instr}.  The instrumentation starts with array length
loading and comparison against the attacker provided index, at
\prettyref{line:rustasm_cmp}. Depending on the comparison outcome, the
execution proceeds with accessing the array element requested or aborting in
case of in-bounds requirement violation.

When the comparison between index and length is slow (due to uncached
operands), the CPU may mispredict the result and continue execution
speculatively, on the wrong path.

In the PoC, the victim data structure is chosen such that the array length can
be evicted prior to the attack. The array length is stored together with the
array data pointer in \Code{buf}.  At \prettyref{line:rustasm_box} the
\Code{Data} object is initialized using \Code{Box}, therefore the object is
placed on heap.  This avoids Rust default stack allocation which lowers the
array length eviction success. Furthermore, the eviction may affect attack
critical data, like the \Code{buf} data pointer. In the PoC, \Code{Data} uses a large
enough padding so that the array length and the data pointer land on different
cache lines.

The \Code{buf} length eviction triggers
mispeculation of the jump direction taken (\prettyref{lst:rust_instr},
\prettyref{line:rustasm_jmp}). Inside the speculation window, an out of bounds
array access with the attacker controlled index leads to reading a function
pointer from an attacker-owned memory area.
Subsequently, the attacker controlled function pointer is the \Code{call}
instruction destination (\prettyref{line:rustasm_call}), therefore facilitating
speculative execution of attacker chosen code (in
this case, a speculation marker). Despite of the CPU rolling back the
speculative execution effects on registers and memory, we use Intel
Performance Monitoring Counters for counting speculation marker hits. We
carry out the experiments on an Intel Skylake machine running Ubuntu 18.04. We
measure an overall success rate of $90$\% ($n{=}1000$) for the \spac attack against Rust bounds
checking mechanism. As for \Go and GCC VTV, this success rate refers to the
hijack phase only.

\section{Further Code Snippets} \label{app:snippets}

\begin{lstlisting}[frame=none,
                   language={[x64]Assembler},
                   caption=Architectural forward edge overwrite.,
                   basicstyle=\scriptsize\ttfamily,
                   label=lst:arc_fwd,
                   escapechar=|]
;Copy of Target Value
    mov rax, [orig_target]
    mov QWORD[stored_target], rax |\label{line:copy_target}|

;Architectural Overwrite
; (Attacker Controlled)
    mov rax, QWORD[hijacked_target]
    mov QWORD[target], rax  |\label{line:overwrite}|

;Evict Target Value Copy
    clflush [stored_target] |\label{line:evict}|
    lfence

;Forward Edge Integrity Check
; (Speculation Trigger)
    mov rax, QWORD[target]
    cmp rax, QWORD[stored_target] |\label{line:intcheck}|
    jne my_exit

;Forward Edge Hijack
    call QWORD[target]      |\label{line:CFH}
\end{lstlisting}

\begin{lstlisting}[frame=none,
                   basicstyle=\scriptsize\ttfamily,
                   language={[x64]Assembler},
                   caption=Speculative forward edge overwrite.,
                   label=lst:spec_fwd]
;Speculative execution trigger
    ...

;Speculative Overwrite
; (Attacker Controlled)
    mov rax, QWORD[hijacked_target]
    mov QWORD[target], rax

;Forward Edge Hijack
    call QWORD[target]
\end{lstlisting}

\end{appendices}

\onecolumn

\end{document}